\documentstyle[11pt]{article}
\renewcommand{\baselinestretch}{1.3}

\input epsf

\begin{document}
\newcount\nummer \nummer=0
\def\f#1{\global\advance\nummer by 1 \eqno{(\number\nummer)}
      \global\edef#1{(\number\nummer)}}

\newcommand{\pref}[1]{\ref{#1}}                      
\newcommand{\plabel}[1]{\label{#1}}                  

\newcommand{\prefeq}[1]{Gl.~(\ref{#1})}              
\newcommand{\prefb}[1]{(\ref{#1})}                   
\newcommand{\prefapp}[1]{Appendix~\ref{#1}}          
\newcommand{\plititem}[1]{\begin{zitat}{#1}\end{zitat}}     
\newcommand{\plookup}[1]{\hoch{\ref{#1}}}

\let\oe=\o
\def\Di{\displaystyle}
\def\nn{\nonumber \\}
\def\be{\begin{equation}}
\def\ee{\end{equation}}
\def\ba{\begin{eqnarray}}
\def\ea{\end{eqnarray}}
\def\la{\plabel} \def\pl{\label}
\def\re{(\ref }
 \def\rz#1 {(\ref{#1}) } \def\ry#1 {(\ref{#1})}
\def\el#1 {\plabel{#1}\end{equation}}
\def\rp#1 {(\ref{#1}) }
\def\i{{\rm i}}
\let\a=\alpha \let\b=\beta \let\g=\gamma \let\d=\delta
\let\e=\varepsilon \let\ep=\epsilon \let\z=\zeta \let\h=\eta \let\th=\theta
\let\dh=\vartheta \let\k=\kappa \let\l=\lambda \let\m=\mu
\let\n=\nu \let\x=\xi \let\p=\pi \let\r=\rho \let\s=\sigma
\let\t=\tau \let\o=\omega \let\c=\chi \let\ps=\psi
\let\ph=\varphi \let\Ph=\phi \let\PH=\Phi \let\Ps=\Psi
\let\O=\Omega \let\S=\Sigma \let\P=\Pi \let\Th=\Theta
\let\L=\Lambda \let\G=\Gamma \let\D=\Delta

\def\wt{\widetilde}
\def\w{\wedge}
\def\0{\over } \def\1{\vec } \def\2{{1\over2}} \def\4{{1\over4}}
\def\5{\bar } \def\6{\partial }
\def\7#1{{#1}\llap{/}}
\def\8#1{{\textstyle{#1}}} \def\9#1{{\bf {#1}}}

\def\({\left(} \def\){\right)} \def\<{\langle } \def\>{\rangle }
\def\lb{\left\{} \def\rb{\right\}}
\let\lra=\leftrightarrow \let\LRA=\Leftrightarrow
\let\Ra=\Rightarrow \let\ra=\rightarrow
\def\ul{\underline}

\let\ap=\approx \let\eq=\equiv 
\let\ti=\tilde \let\bl=\biggl \let\br=\biggr
\let\bi=\choose \let\at=\atop \let\mat=\pmatrix
\def\CL{{\cal L}}\def\CX{{\cal X}}\def\CA{{\cal A}}
\def\CF{{\cal F}} \def\CD{{\cal D}} \def\rd{{\rm d}} 
\def\rD{{\rm D}} \def\CH{{\cal H}} \def\CT{{\cal T}} \def\CM{{\cal M}}
\def\CI{{\cal I}} \newcommand{\dR}{\mbox{{\sl I \hspace{-0.8em} R}}} 
  \newcommand{\dN}{\mbox{{\sl I \hspace{-0.8em} N}}}
\def\CP{{\cal P}}\def\CS{{\cal S}}\def\C{{\cal C}}

\begin{titlepage}
\renewcommand{\thefootnote}{\fnsymbol{footnote}}
\renewcommand{\baselinestretch}{1.3}
\hfill  TUW - 95 - 13 \\
\medskip
\hfill  PITHA - 95/10 \\
\medskip
\hfill  gr-qc/9507011 \\
\medskip

\begin{center}
{\LARGE {Explicit Global Coordinates for Schwarzschild and
Reissner-Nordstr\oe m}}
\medskip 
\vfill
\renewcommand{\baselinestretch}{1} 
{\large {THOMAS 
KL\"OSCH}\footnote{e-mail: kloesch@tph.tuwien.ac.at\newline
\hspace*{8 pt}
$^\diamondsuit$e-mail: tstrobl@pluto.physik.rwth-aachen.de} \\
\medskip 
Institut f\"ur Theoretische Physik \\
Technische Universit\"at Wien\\
Wiedner Hauptstr. 8-10, A-1040 Vienna\\
Austria\\ }
\medskip 
\medskip {\large THOMAS STROBL$^\diamondsuit$
 \\ \medskip
Institut f\"ur Theoretische Physik \\
RWTH-Aachen\\
Sommerfeldstr. 26-28, D-52056 Aachen \\
 Germany \\ } 
\end{center}

\vfill
\renewcommand{\baselinestretch}{1}                          

\begin{abstract}
We construct coordinate systems that cover all of the 
Reissner-Nordstr\oe m solution with  $m > |q|$ and $m=|q|$,
respectively. This is possible by means 
of elementary analytical functions. The limit of vanishing charge 
$q$ provides an alternative to Kruskal which, to our mind,   
is more explicit and  simpler. 
The main tool for finding these global charts is the  
description of highly symmetrical metrics by two-dimensional actions.
Careful gauge fixing yields global representatives of the 
two-dimensional theory that can be rewritten easily as 
the corresponding four-dimensional line elements.
\end{abstract}  

\vfill
\noindent to appear in {\em Class.\ Quantum Grav.}\hfill June 1995\\ 
\end{titlepage}

\setcounter{footnote}{0}

The purpose of this letter is twofold: First we want to  
present the Reissner-Nordstr\oe m (RN) and the Schwarzschild (SS)
solution within one global coordinate system.  It is remarkable that in
both cases this is possible by means of elementary 
functions only. For the SS metric there 
certainly exist already the global Kruskal-Szekeres coordinates
\cite{Krus}, in which 
\be ds^2= {32 m^3 \0 r} e^{-r/2m} (dT^2-dX^2)
- r^2 d\O^2 \, \el Kruskal
where $r\equiv r(X,T)$ is defined {\em implicitly} by 
  $[(r/2m)-1]\exp(r/2m)=X^2-T^2$ and $d\O^2 = (d\vartheta^2 + 
\sin^2\vartheta d \varphi^2)$ denotes the standard metric on the 
two-sphere. Our  SS line element, on the other
  hand,  has the simple explicit form (cf.\ also Fig.\ 1) 
\be ds^2 = -8m\,\left[ dxdy + {\, y^2 \0 xy +2m} dx^2 \right] -
(xy +2m)^2  d\O^2 \, . \el SS
Also the transformation to the standard SS-form is straightforward 
in both directions (cf.\ Eqs.\ \rz to and \rz back below).
Our  line element for RN will be not that concise, but still it
contains only nonsingular ratios of 
trigonometric functions and provides an elementary description of RN within
one analytical chart (cf.\ also Fig.\ 2).

Certainly  in the first place such  global coordinate systems 
 will be of educational importance. 
Still it is to be expected that also one or the other calculation 
of relevance for modern research will be facilitated by the use of 
simple global coordinates.  

Our second interest, related to the first one as 
outlined below, shall 
be  an investigation of two-dimensional gravity theories,  
more precisely, of (reformulated) generalized dilaton gravities 
\cite{Kunst, Banks} 
\be S[g,\phi]= -\mbox{$\2$}
\int_M d^2x \sqrt{-\det g} \,\, [\phi R - V(\phi)]  \, . \el Dil
Here $g=g_{\m\n} dx^\m dx^\n$ is understood to be a 
Minkowskian metric on a 
{\em two}-manifold $M$ with coordinates $x^\m$, $\m = 0,1$, $R$ is the 
corresponding Levi-Civita curvature scalar, $\phi$ is a 
function on $M$ (the dilaton field), 
and $V$ is some given smooth potential.  
Using an Einstein-Cartan formulation, \rz Dil may be rewritten into 
\be S[e^a,\o,\varphi_a,\phi] = \int_M \varphi_a \, De^a + \phi \, d\o 
-  \mbox{$\2$}V(\phi) \, e^+ \wedge e^- \, , \el Pal 
where $e^a \, , a \in \{ +,- \}$, is the {\em zweibein}, $g=2e^+e^-$, 
and $\varepsilon^a{}_b \,  \o$ is the  {\em spin}-connection\footnote{In our
 conventions $\varepsilon^{+-}=1$
and, e.g., $\varphi_\pm = \varphi^\mp$.}. 
(The equivalence between \rz Pal and \rz Dil becomes 
obvious when realizing 
that the variation with respect to the Lagrange multiplier fields 
$\varphi_a$ yields just the torsion zero condition $De^a \equiv 
de^a + \varepsilon^a{}_b \o \wedge e^b=0$; 
the remaining part of the action then 
coincides with \rz Dil since $R= 2\ast d\o$, where ``$\ast$'' denotes the
Hodge dual).  

In \cite{Peter} and, more explicitly, in \cite{Klo} 
it is shown that a model of the type \rz Pal may be solved very efficiently
when reinterpreting it as a  Poisson $\sigma$-model, i.e.\ a $\sigma$-model
where the target space $N$ is a  Poisson manifold. The particular action
\rz Pal turns out to correspond to a target space $N=\dR^3$ carrying the
Poisson bracket 
$\{\varphi^+,\varphi^-\}=\mbox{$\2$}V(\phi)$, $\{\varphi^a, \phi\}
=\varepsilon^a{}_b \varphi^b$. With the use of Poisson structure
adapted coordinates on $N$, the equations of 
motion (e.o.m.) may be solved locally in a couple of lines of calculation.
It is shown here that within the more elementary approach of gauge
fixation the solution to the field equations may be immediate as well. 
And for the purpose of constructing solutions in explicit form 
that have global validity {\em without} patching (if existent),    
the present method proves even preferable. 

The relation between the two and the four dimensions comes in as
follows. It is well-known that locally the RN-solution may be characterized by
\ba ds^2&=&h_{RN}(r) dt^2 - {1 \0 h_{RN}(r)} dr^2 -r^2 d\O^2 \label{gRN} \\
h_{RN}(r) &=&1- {2m \0 r} + {q^2 \0 r^2}  \, . \label{hRN} \ea
On the other hand, any local solution to the field equations of 
\rz Dil is of the form \cite{thesis,Kunstneu,Klo} 
\be g=h(r) dt^2 - {1 \0 h(r)} dr^2 \, , \quad  \quad \phi =r \, ,\el glocal
with 
\be h(r)=\int^r V(u)du + const. \el h
(where the indetermination in the integration of \rz h gives 
rise to a one-parameter family of distinct solutions). 
So we may conclude that \rz Pal with the potential
\be V^{RN} = h_{RN}'(\phi) \equiv 
{2m \0 \phi^2} - {2q^2 \0 \phi^3 }  \,  \el VRN 
can be taken to describe RN (respectively SS for $q=0$),  
if the four-dimensional metric $g_{(4)}$ is determined via  
\be g_{(4)} = g -\phi^2 d\O^2 \, . \el g4
 
The strategy of this letter is the following: We  pick some gauge that may be
enforced {\em globally} on space-times of a sufficiently large class so as to
cover RN and SS. In this gauge we  solve the e.o.m.\ of \rz Dil or \ry Pal .
As a first exercise we recover the local result \ry glocal , or 
\ry gRN , respectively. Thereafter, however, we carefully investigate the
residual gauge freedom of the chosen gauge so as to obtain global charts. 
These charts are then applied to the specific cases of SS and RN.
 
To break the diffeomorphism invariance we choose the light cone gauge  
\be g= 2dx^0dx^1 + k(x^0,x^1) (dx^1)^2 \, . \el gauge1
The region of validity of \rz gauge1 can be found as follows: First 
we label one set of null lines on $M$ by the $x^1$-coordinate. This implies
$g_{00}=0$. Then we choose the second coordinate 
$x^0$ to coincide with some affine parameter 
along these null extremals. This may be seen to yield $g_{01}$  
independent of $x^0$. As an affine parameter is determined only up to
linear transformations, which in our case may be 
$x^1$-dependent,  one may absorb $g_{01}(x^1)$ into $x^0$ 
by an appropriate rescaling of the affine parameter.  
The result is \ry gauge1 . Obviously this procedure works on any part
of the two-dimensional space-time $M$ that may be foliated into
null-like lines. Thus in particular it will work all over  (two- or 
four-dimensional) RN and SS space-time (cf.\ also Fig.\ 2a). 

Describing the metric \rz gauge1 by a {\em zweibein}, 
clearly one of the two one-forms $e^\pm$ 
has to have a vanishing coefficient function in front of 
$dx^0$. A globally attainable choice of the Lorentz frame 
then brings the {\em zweibein} into  the form:
\be e^+ = dx^1 \; , \quad e^-=dx^0 + \mbox{$\2$} k\, dx^1  \, . \el gauge2

Let us now determine the function $k$ in the gauge \rz gauge2 by means
of the e.o.m.\  of \ry Pal . Beside torsion zero and
$R=V'(\phi)$ the e.o.m.\ are
\ba d\phi + \varepsilon_{ab}\varphi^a e^b &=& 0 \, \nn 
d\varphi_a + \varepsilon_{ab} \varphi^b \o + \mbox{$\2$}
\varepsilon_{ab} \, V(\phi) \, e^b &=&0 \label{var} \, . \ea
{}From $De^+=0$ we learn at once that
$\o_0=0$. Then the $x^0$-components of 
\rz var (with $a=-$) become  $\6_0 \varphi^+ = 0$ and
$\6_0 \phi = \varphi^+$,  respectively. This yields 
\be \varphi^+=F(x^1) \; , \quad \phi= F(x^1) x^0 + G(x^1) \, , \el Loes
for some functions $F$ and $G$. Next, the Eqs.\ \rz var may be seen to imply
$V(\phi)d\phi= d(\varphi^2)$ where $\varphi^2 \equiv \varphi^a\varphi_a$. So
\be \varphi^2 \equiv 2 \varphi^+\varphi^- = h(\phi) \, ,
\el phi2
where the function $h$ has been defined in \rz h already. Writing out the 
one-components of the field equations \rz var with ($a=-$), 
\ba \6_1 \varphi^+ + \mbox{$\2$} V + \varphi^+ \o_1 &=&0 \, , \label{G+} \\
\6_1 \phi + \varphi^- - \mbox{$\2$}  \varphi^+ k &=& 0 \, , \label{G3}
\ea
we realize that the second of these equations determines 
$k$ in terms of $F$, $G$, and the integration constant of \re{phi2}),\re{h}): 
\be k= 2 {F' x^0 + G' \0 F}+ {h(Fx^0 +G) \0 F^2} \, . \el k

The functions $F$ and $G$ are not completely unrestricted, however. 
{}From \rz phi2 and \rz Loes we 
infer \be h\left( G|_{F=0} \right) = 0  \, . \el restrict2
Similarly \rz G+ yields
\be F'|_{F=0} = -\mbox{$\2$} V|_{F=0} \equiv -\mbox{$\2$} h'\left(
G|_{F=0} \right) \, , \el restrict1 
where we used \rz h  for the second equality. 
In the case of simple zeros of $F$ (sufficient for analyzing SS and 
non-extreme RN) it is straightforward to verify that these 
restrictions suffice to render $k$ smooth for smooth choices of  $F$ and 
$G$.  \rz restrict2 and \rz restrict1 are, moreover, the {\em only} 
restrictions on $F$ and $G$ beside smoothness. This will become 
obvious when analyzing the residual gauge freedom left by \rz gauge1 and
\ry gauge2 . Indeed it will be found below that, for a fixed number of
(simple) zeros of $F$, any set of functions $F$ and $G$ respecting  
\rz restrict2 and \rz restrict1 are related to each other by gauge 
transformations.

On parts of $M$ that may be foliated by null-extremals, (\ref{gauge1},
\ref{k}--\ref{restrict1}) provides the general solution of the field 
equations of \rz Dil or \rz Pal already 
(parametrized by the integration constant in \rz h and the number 
of zeros of $F$).  Let us mention here that the gauge \re{gauge2}) 
works similarly efficient, if one allows $V$ in \rz Pal to depend on
$\varphi^2$, too. Such an action yields solutions with nontrivial torsion
then. Also, similarly one might have used the (``Hamiltonian'') gauge 
$e_0^-=1, e_0^+=0=\o_0$ \cite{Kummer}, which is attainable globally as well,
if $M$ can be foliated by null-lines. The use of a conformal gauge seems less
advisable here, however. 

We now turn to an analysis of the residual gauge freedom of our gauge 
conditions: First,  \rz gauge1 remains unaffected, if an $x^1$-dependent
linear transformation of the affine parameter $x^0$ is compensated by a
diffeomorphism in the $x^1$-variable: 
\ba x^0&=& {1 \0 f'(\wt x^1)} \wt x^0 + l(\wt x^1)  \; , \nn
x^1&=& f(\wt x^1)  \;, \quad  f'(\wt x^1) \neq 0 \; . \label{diffeo} \ea
An additional Lorentz transformation 
\be e^- \ra  f'(\wt x^1) \, e^- \; , \quad e^+ \ra {1 \0 f'(\wt x^1)} \, e^+
\el Loren
is necessary to restore  \ry gauge2 . 
For $F$ and $G$ in \rz Loes this 
implies the following equivalence relations:
\be F(x) \sim {F(f(x))\0 f'(x)}\; , \quad G(x) 
\sim G(f(x)) + F(f(x)) \,  l(x) \; .
\label{equiv} \ee 

As a warm up in the study of \rz equiv let us first 
consider local patches on $M$ with $\varphi^+ =F \neq 0$ (in which case
\rz restrict2 and \rz restrict1 are  empty). 
On such a patch $F \sim  1$, because the 
differential equation $f'(x)=F(f(x))$ may  be solved for 
a monotonous $f$. The subsequent choice $f(x)=x$ and $l(x)=-G(x)$ in 
\rz equiv yields $G \sim 0$, furthermore. 
So on patches with $\varphi^+ \neq 0$ we may put $\varphi^+=1$ 
and $\phi=x^0$ by residual gauge
transformations  ($\leftrightarrow \, F=1 \, , \; G=0$). 
Then \rz k reduces to 
\be k(x^0,x^1)=h(x^0) \quad \mbox{(locally)} \, . \el Finkel

Thus locally the metric $g$ takes the generalized
Eddington-Finkelstein form  \re{g4}, \ref{gauge1}, \ref{Finkel}).  
This form is particularly well-suited for the construction of Penrose
diagrams (cf., e.g., \cite{Klo}).  It breaks down at zeros of
$\varphi^+$. On patches restricted further by $\varphi^- \neq
0$,  the diffeomorphism
\be r:= x^0 \; , \quad t:= x^1 + \int^{x^0}\!\!\!{dz \0 h(z) } \el SSdiff
  brings $g$ into the generalized Schwarzschild form \rz glocal 
(with  $h$ unmodified), thus confirming the statement on local
solutions made in the introduction.  The integration constant in
$h$ is then left as the only locally meaningful parameter that
cannot be gauged away.  (In the case of
\rz VRN this parameter may be seen to effectively rescale $m$
and $q$).  The range of such patches within a Penrose diagram
becomes clear when noting that according to \rz phi2 Killing
horizons are labeled by zeros of $\varphi^\pm$, cf.\ Figs.\ 2.

Let us now discuss \rz equiv in the general setting of zeros of
$F$, restricting ourselves at a first stage, however,  
to the case of simple zeros of $F$. Maybe it is worth mentioning 
that solving the first relation 
 \rz equiv is equivalent to a classification of 
all $C^\infty$-vectorfields $v:=F(x) d/dx$  on a real line up 
to diffeomorphisms. One finds 
that  the zeros of $F$ may be shifted along the $x \sim x^1$-coordinate line
at will, while their total number cannot be changed.\footnote{Actually
for $x \in \dR$
there exist also coordinate transformations that boost zeros of $F$ into
infinity. This, however,  leads to coordinate singularities at
``infinity''.} Also it is  not difficult to verify 
that the linear coefficients $F'|_{F=0}$ 
may not be changed by the symmetry relations \ry equiv . For 
the case of 
an open interval for the variable $x \sim x^1$ this is all that remains for
$F$: The number of its zeros  together with the corresponding 
slopes at those points. For the case of simple zeros of $F$ the second
equivalence relation \rz equiv may be solved readily also: $G$ may be
``deformed'' in an arbitrary way except for those points where $F$
vanishes; there the value of $G$ remains fixed.

All the features of $F$ and $G$ that may not be changed by the
gauge transformations \rz equiv  are determined already  by \rz restrict1 
and \rz restrict2 (except for the number of zeros of $F$ and
possibly a further discrete choice in solving  \ry restrict2 ).
For means of completeness let us mention here also that in the
case of a closed line $x \sim x^1$, describing 2D-gravity
solutions to the field equations with cylindrical topology,
there remains one further unrestricted parameter in the analysis
of \ry equiv . The appearance of a second continuous parameter 
(beside the integration constant hidden in $h$) for cylindrical
space-time topology is important in the context of a Hamiltonian
analysis of the theory \rz Pal  (or \ry Dil ). It reveals that
the reduced phase space will be two-dimensional;  the conservation of 
zeros of $F$ indicates nontrivial topology of this phase space. 
More details on such aspects may be found in \cite{Klo}, \cite{thesis}.

Picking representatives $F$ and $G$ in the global setting, where
zeros of $F$ are not excluded, we will restrict ourselves to the SS and RN
case \rz VRN in the following. Also the integration constant in \rz h 
shall be chosen such that $h$ coincides with $h_{RN}$. The generalization 
 to arbitrary $h$ will be obvious, however. We begin with the simpler case 
$q=0$ describing SS: Here 
$h$ has  one zero at $2m$. Thus \rz restrict2 implies 
$G|_{F=0}=2m$ and \rz restrict1 $F'|_{F=0}=-1/4m <0$. As a consequence $F$ may
have one zero only. The  residual gauge transformations then allow to set 
\be F(x^1) = -{x^1 \0 4m } \; , \quad G(x^1)=2m \, . \el SSFG
Insertion into \rz k yields 
\be k = {2(x^0)^2 \0 x^0x^1 - 8m^2 } \, , \el kSS
or, as $\phi =2m -x^1x^0/4m$, the four-dimensional line element
\rz SS after the  rescaling $x:=x^1$, $y:=-x^0/4m$. 

Maybe it is worth mentioning that by a coordinate rescaling the mass
parameter of SS (or also of RN) may be  separated into an overall conformal
factor. Here this implies that instead of \rz SS we may write also 
\be ds^2= -4m^2 \left [ 4 d \widetilde  x d \widetilde y + {4 \widetilde y^2
\0
\widetilde x \widetilde y+1 } d \widetilde x^2  + ( \widetilde x
\widetilde y+1)^2
 d \Omega^2  \right] \, . \ee
                       
Certainly it is  not difficult to find the local coordinate
transformation that maps \rz SS into the standard
Schwarzschild form \ry gRN .
First the transformation\footnote{It is found most
easily when noting that, for $x \neq 0$,   \rz Finkel is obtained 
{}from transforming  $F$ into 1 and $\phi$ into $x^0$ ($\lra G=0$). The
first equation of \rz Trafo is then obvious and the second one results
from $x=f(x^1)$ and $F(f(x^1)) := f'(x^1)$,
cf.\ Eqs.\ (\ref{diffeo}, \ref{equiv}). (The coordinates $x^\m$ in
\re{Trafo}) are, of course, different from those in \re{SSFG},\ref{kSS})!)}
\be x^0 := \phi = xy+2m \, , \quad x^1 := -4m \, \ln|x| \el Trafo
brings \rz SS into Eddington-Finkelstein form \ry Finkel . The  subsequent
transformation \ry SSdiff , which becomes 
$r:=x^0$, $t:= x^1 + x^0 + 2m \, \ln |x^0 -2m|$  here, yields the 
Schwarzschild form. Putting these transformations  together, 
one has 
\ba r &=& xy + 2m \nn
t &=& xy + 2m(1+\ln|y/x|)  \, . \label{to} \ea 
So, obviously not only the resulting line element \ry SS , 
but also the transformation from the standard SS-form 
to it, has a particularly simple form. Inverting \ry to , one finds 
\ba 
|x| &=& \sqrt{|r-2m|} \exp \left({r-t \0 4m} \right) \nn
|y| &=&\sqrt{|r-2m|} \exp \left({t-r \0 4m} \right) \, .\label{back} \ea 

It is straightforward to determine the radial null-directions of
\re{SS}). 
They are $x=z_1$ and $y \exp(xy/2m+1) = 8m \,  z_2$, respectively, with
$z_1$, $z_2$ constant. Clearly, if we introduced $z_1$ and $z_2$, or,
more precisely, $X=(z_1 + z_2)$ and $T=(z_1-z_2)$, as new coordinates, 
the metric \rz SS would  be brought into the Kruskal-Szekeres form \ry
Kruskal . However, the causal structure is already captured completely in
the $x,y$-coordinates, cf.\ Fig.~1.

Let us turn to the RN case $q \neq 0$ with $m>|q|$ next. In this case  
$h(r)=h_{RN}(r)$ has two simple zeros  at 
\be r_\pm = m \pm \sqrt{m^2 - q^2} \, . \el r+-
Labelling the zeros of $F$ by $x^1 = n \pi$, $n \in {\bf Z}$, $F$ has the form
\be F(x^1)= \a(x^1) \sin x^1 \el F
with $\a \neq 0$. Then 
\rz restrict2 becomes $G(n\pi) \in \{ r_-, r_+ \}$, which is respected, if
we choose, e.g.,  $G={ r_+ + r_- \0 2} - { r_+ - r_- \0 2} \cos x^1$, i.e.\ 
\be G(x^1)= m  - \sqrt{ m^2 -q^2} \cos x^1 \, . \el G
The restriction \ry restrict1 , on the other hand, is one for $\a$ now:
$\a(n\pi)= - \mbox{$\2$} h_{RN}'(r_-)=:\a_-$ for $n$ even and $\a(n\pi)=
 \mbox{$\2$} h_{RN}'(r_+)=:\a_+$ for $n$ odd. 
A simple possible choice for $\a$ is  
\be \a(x^1)= C_1 + C_2 \cos x^1  \el alpha
with  
$C_{1,2}={\a_- \pm \a_+\0 2}$, i.e.\ with
$C_1 = \sqrt{m^2- q^2}(2m^2-q^2)/q^4$ and $C_2 = (m^2-q^2)\,  2m/q^4$.
The full four-dimensional RN metric is then given by  
\ba ds^2 &= &2dx^0dx^1+\frac{2\partial_1 \phi +h_{RN}(\phi)/F}F (dx^1)^2
      - \phi^2 d\Omega^2 \, , \nn
\phi&=&F(x^1)x^0 + G(x^1) \, , 
\label{gRNfull} \ea 
with the functions $F$ and $G$ as above and $h_{RN}$ as in \ry hRN . The
result is depicted in Fig.\ 2a. 

The coefficient function $k$ in front of the $(dx^1)^2$-term of \rz gRNfull is
manifestly analytic all over $M$, 
as, by construction, it is a non-singular quotient of analytical functions
only. Actually, it is instructive to regard $k$  in the vicinity of the
saddle points of $\phi$ at $r_\pm$ (where $F=0$). 
For this purpose we replace  $F$ and $G$ in  \rz G and \rz F by their linear
approximations close to the zeros of $F$: 
\be F= \mp\a_\pm x^1 \; , \quad G= r_\pm  \, . \el linear 
With such a choice $k$ takes the form
\be k=\frac{2{\a_\pm}^2 x^0x^1 \mp 4 r_\pm \a_\pm +1}{(\mp \a_\pm x^0x^1 +
r_\pm)^2}(x^0)^2  \, ,  \el saddle
generalizing the chart \rz SS for Schwarzschild. Indeed,
$r_+|_{q=0}=2m$, $\a_+|_{q=0}=1/4m$ so that in the
``+''-chart of \rz saddle $k$ is seen to reduce to \rz kSS for
$q=0$.\footnote{Certainly \rz kSS may be obtained also directly
{}from (\ref{G}, \ref{F})  by an appropriate  limiting procedure $q \to
0$.  One only has to rescale $x^1$ and $x^0$ by $q$ and $1/q$,
respectively, which also  leads to a multiplication 
of $F$ by $1/q$ (cf.\ the first Eq.\ \ry equiv ). One obtains 
 $\lim_{q \to 0} G(qx^1)= 
\lim_{q \to 0}  r_+ = 2m$ and $\lim_{q \to 0} F(qx^1)/q= \lim_{q \to 0}
F'(qx^1) x^1 = -\lim_{q \to 0} \a_+ x^1 = - x^1/4m$, in coincidence 
with \ry SSFG .} 
The representation \rz saddle together with $\phi=  \mp\a_\pm x^0 x^1
+ r_\pm$  may be used to describe RN within a maximal region containing
one zero of $\varphi^+=F(x^1)$, cf.\ Fig.\ 2b. 

Extreme RN, characterized by $m=|q|$, does not follow from (\ref{G},
\ref{F}) as a limit. One of the reasons for this is that we required the 
(coordinate-)distance between two adjacent simple zeros of $F$ to be $\pi$.
In the case of extreme RN $F$ has {\em two}-fold zeros, however: 
$F'|_{F=0}=0$,  cf.\ Eq.\ \ry restrict1 . This may be described
successfully by a limiting procedure $m \to |q|$ only, 
if the (coordinate-)distance between two adjacent
zeros of $F$ shrinks with $m \to |q|$, while the distance between such
pairs of zeros of $F$ remains finite. But there are also further 
restrictions for a finite result 
(cf., e.g., Eq.\ \rz restrict3 below). Thus, for simplicity, we
treat extreme RN by our gauge fixing method as an independent case.

As $F$ has a two-fold zero now, the analysis of \rz equiv becomes more
involved. Still it is straightforward to see that 
in such a case there are two further quantities at each zero of $F$ 
which are invariant under the transformations \ry equiv , namely, the 
ratios between $G'$ and $F''$ as well as between $F'''$ and $(F'')^2$. 
Again these quantities are fixed by the equations of motion and 
everything else is pure gauge (except certainly, as before, 
 $G|_{F=0}$, $F'|_{F=0}$, and the number of zeros of $F$).  
E.g.\ differentiation of \rz G+ with respect to  $x^1$ yields 
$2F''|_{F=0}=- (V'G')|_{F=0}$, or, when labeling the zeros of $F$ again by
$x^1=n\pi$
\be F''(n\pi)=-{G'(n\pi) \0 m^2} \, , \el restrict3
where we made use of the fact that $G(n\pi)=m$ and $V'(m)=h_{RN}''(m)=2/m^2$. 
For extreme RN a possible 
choice for $F$ and $G$ is 
\ba F(x)&=& \frac{\cos x-1}{2m}\left(\frac{\sin x}2-1\right)^2 \nn
  G(x)&=& m \left(\frac{\sin x}2+1\right)  \, .  \ea
Extreme RN has been presented in a global chart already in \cite{Carter},
but again the functions involved in the description were defined implicitly
only.   

Concluding, it seems remarkable to us that two-dimensional
gravity models proved to be a powerful tool for the construction
of global charts for {\em four}-dimensional
Reissner-Nordstr\oe m and Schwarzschild space-time. We started with an
appropriate 2D Lagrangian and  picked globally attainable gauge conditions.
They allowed us to solve the field equations on a global level. 
Exploiting the residual gauge freedom on-shell,
the global representatives of SS and RN, presented in this letter, 
just popped up.

The models \rz Pal are studied further in their own right
in \cite{Klo} (but cf.\ also \cite{Kunst, Peter, thesis, Kunstneu,
Kunstviele, LNP}).

\section*{Acknowledgement}
We are grateful to H.\ Balasin, H.D.\ Conradi, and W.\ Kummer
for  encouraging us to write this letter and for discussions.
Also we profited from discussions  with Peter Schaller
on gauge fixing within one or the other 2D gravity-model.
This work has been supported in part by the Austrian Fonds zur F\"orderung
der wissenschaftlichen Forschung (Project P10221-PHY).

\section*{Note added}

After completion of this work, J.N. Goldberg informed us that the chart
\re{SS}) for Schwarzschild has been constructed already in
\cite{Newman, Israel}. W. Israel \cite{Israel}, furthermore, gives a chart
for Reissner-Nordstr{\oe}m which is equivalent to our chart
\re{g4},\ref{gauge1},\ref{saddle}). However, in contrast to our other chart
(\ref{F}--\ref{gRNfull}), the one given in \cite{Israel} does not quite cover
the whole manifold, cf.\ Figs.\ 2a,b.

\begin{samepage}
\section*{Figures}
\begin{center}
\leavevmode
\epsfxsize \textwidth \epsfbox{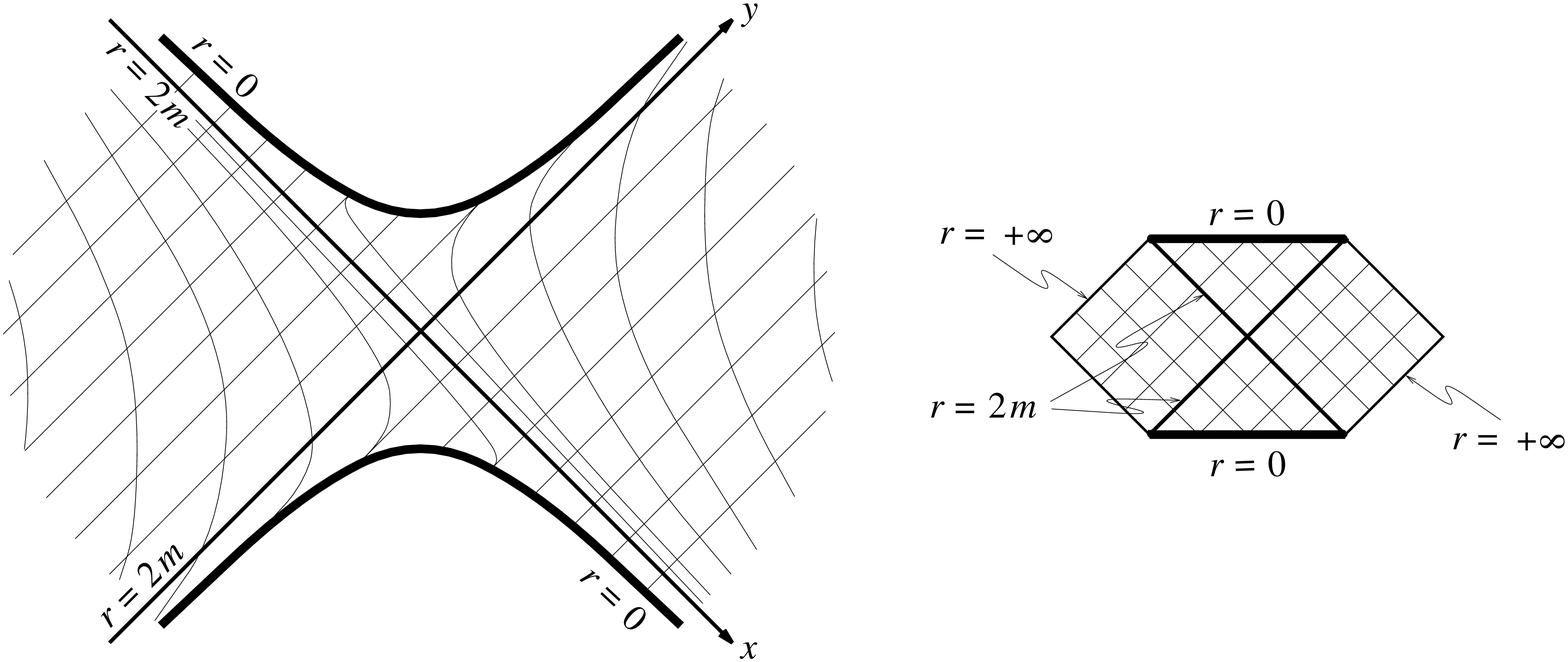}
\end{center}
\renewcommand{\baselinestretch}{1}
\small \normalsize
{\bf Figure 1:} {\small The Schwarzschild space-time. Left coordinates
\re{SS}), right the Penrose diagram. Thin lines represent null-extremals.}
\end{samepage}
\vfill
\pagebreak
\begin{figure}
\begin{center}
\leavevmode
\epsfxsize \textwidth \epsfbox{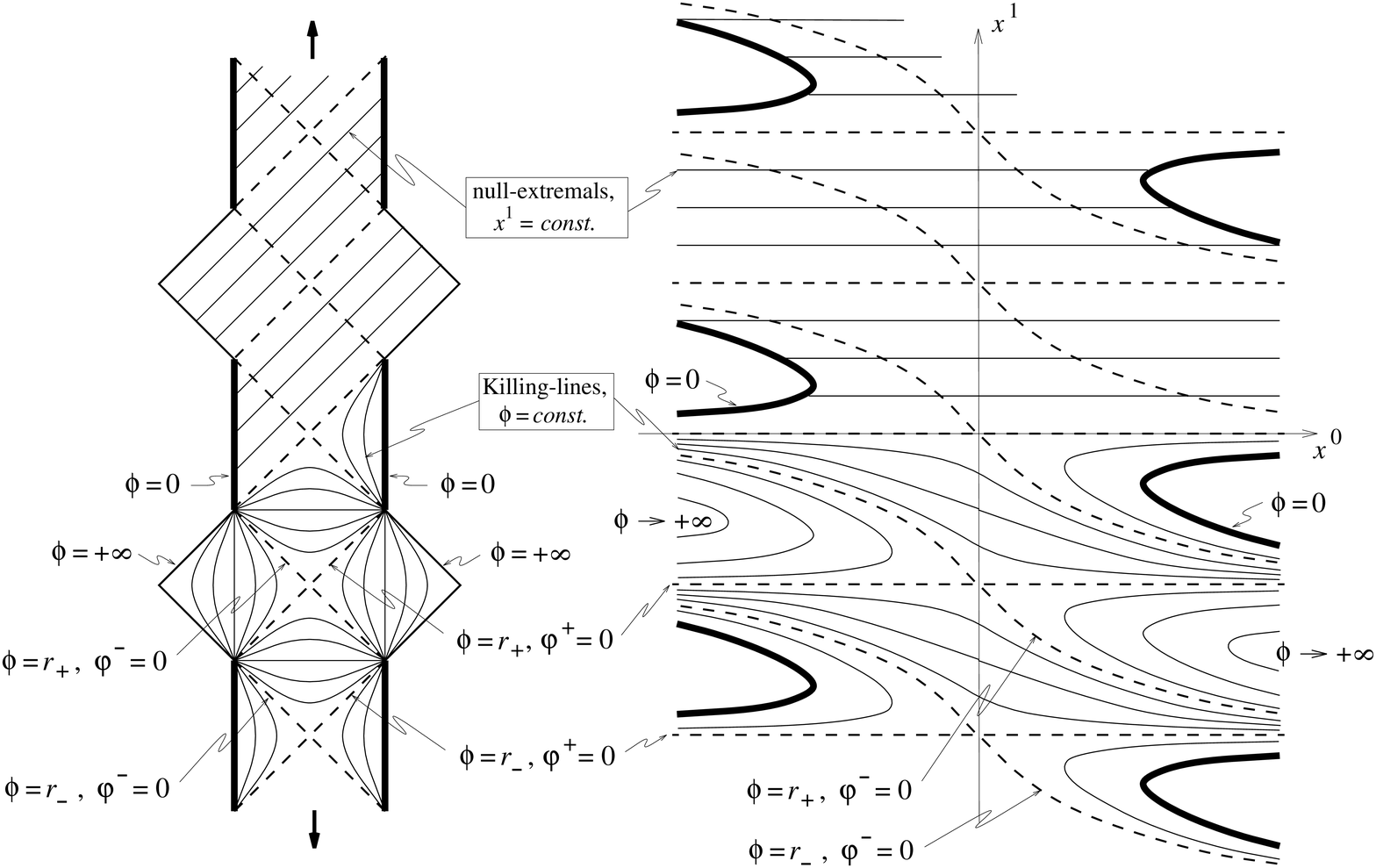}
\end{center}
\renewcommand{\baselinestretch}{.9}
\small \normalsize
{\bf Figure 2a:} {\small Maximally extended Reissner-Nordstr\oe m
space-time for $m>|q|$. Left the Penrose diagram (to
be continued periodically), right the global coordinate system
(\ref{F}--\ref{gRNfull}). In the upper halves we have drawn the
null-extremals $x^1=const$, in the lower
halves the Killing-trajectories $r\equiv\phi=const$.
The usefulness of the auxiliary fields
$\phi$ and $\varphi^\pm$, present in the Lagrangian \ry Pal ,  is obvious:
While $\phi$ coincides with the radius function $r$,  the (Killing) horizons
are marked by the zeros of $\varphi^+=F(x^1)$ and $\varphi^-$.}
\end{figure}
\begin{figure}
\begin{center}
\leavevmode
\epsfysize 6.5cm \epsfbox{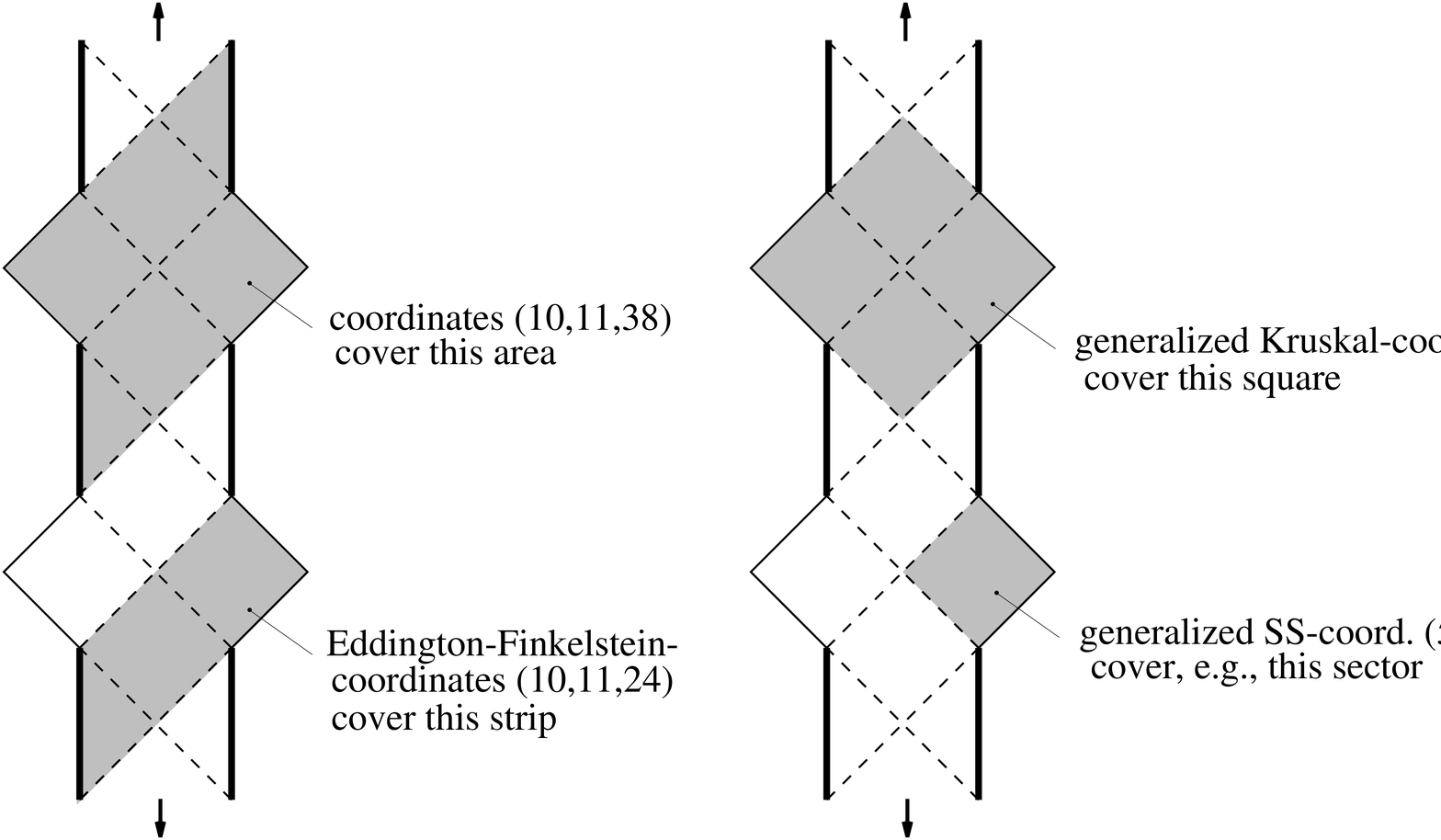}
\end{center}
\renewcommand{\baselinestretch}{1}
\small \normalsize
{\bf Figure 2b:} {\small Range of the different coordinate systems for RN.
Coordinates (\ref{F}--\ref{gRNfull}) cover, of course, the whole manifold
(cf.\ Fig.\ 2a).}
\end{figure}                     
\end{document}